\documentstyle[epsf]{l-aa}     
\begin{document}
   \thesaurus{11.07.1; 11.04.1; 12.04.3; 04.01.1; 03.13.2; 03.13.6}
   \title{Hubble constant from sosie galaxies and HIPPARCOS geometrical calibration}
 
 
\author{
G. Paturel \inst{1} \and
P. Lanoix \inst{1} \and
P. Teerikorpi  \inst{2} \and
G. Theureau \inst{3+4} \and
L. Bottinelli \inst{3+5} \and
L. Gouguenheim \inst{3+5} \and
N. Renaud \inst{1} \and
O. Witasse \inst{1+6}
}

    \offprints{G. Paturel}
 
   \institute{
              CRAL-Observatoire de Lyon, UMR 5574\\
              F69230 Saint-Genis Laval, FRANCE,\\
\and
              Tuorla Observatory\\ 
              SF-21500 Piikkiö, FINLAND\\
\and
              Observatoire de Paris-Meudon, URA 1757\\
              F92195 Meudon Principal Cedex\\
\and
              Osservatorio di Capodimonte\\
              via Moiariello 16, 80131 Napoli, ITALY\\
\and
              Universit\'e Paris-Sud\\
              F91405 Orsay, FRANCE\\
\and
              Université Claude-Bernard LyonI\\
              F69622 Villeurbanne Cedex\\
             }
 
  \date{Received: April, 1997; Accepted: September, 1998}
 
   \maketitle
 
   \begin{abstract}
New distances, larger than previous ones,
have been obtained for M 31 and M 81 based on the geometrical 
zero-point of the Cepheid Period-luminosity relation provided by the
HIPPARCOS satellite. By combining them with independent determinations we define 
reasonable ranges for the distances of these important calibrating galaxies.

On this basis, we determine the Hubble constant from the method of sosies 
(look-alike) galaxies, galaxies having the same characteristics than the calibrators. 
The method is quite secure because it is purely differential and it
does not depend on any assumption (apart from the natural one that two sosies galaxies
have similar absolute luminosities).
Nevertheless, the Malmquist bias has to be taken into account. The observations
behave exactly as predicted from the analytical formulation of the bias.
Thus, rejecting galaxies which are affected by the
Malmquist bias we derive the Hubble constant:
\begin{eqnarray}
H_o = 60\pm 10 (external)  km.s^{-1}.Mpc^{-1}
\nonumber
\end{eqnarray}
If we strictly use the calibration obtained with HIPPARCOS and if the 
bias found in the Period-Luminosity Relation is considered,
the Hubble constant is smaller than this ($\approx 55 km.s^{-1}.Mpc^{-1}$). 
This gives arguments in favour of the long-distance scale.
We briefly discuss  possible improvements aiming at still reducing the uncertainty.

   \end{abstract}
\section{Introduction}
From recent measurements with HIPPARCOS astrometric satellite
we have obtained new distances for 17 nearby galaxies on the basis of
Cepheid geometrical parallaxes (Paturel et al., 1997). 
This provides us with a new calibration of the extragalactic distance scale. 
It would be interesting to look for the Hubble constant
derived directly from these calibrating
galaxies. Unfortunately, their radial velocities are dominated by 
the local velocity field and not by the cosmological velocity. 
Thus, they cannot be used directly, but can be used for the calibration of a long-range criterion,
like the Tully-Fisher relation (Tully and Fisher, 1977).

The TF relation - relation between the 21-cm line width and the absolute magnitude -
is thus far the best way to extend extragalactic distance measurements beyond the
local universe where calibrators can be measured.
The distance modulus derived from this relation can be written as:
\begin{equation}
\mu = B_{T}^{c} - a.logV_M + b = B_{T}^{c} - a.log(W/2sini) + b
\end{equation}
where $\mu$ is the distance modulus, $B_{T}^{c}$ is the apparent total magnitude
corrected for inclination and galactic extinction, $V_M$ is the maximum velocity rotation,
$W$ is the 21-cm line width corrected for instrumental effect and non circular
velocity, $a$ and $b$ are two constants. 
This seems quite simple but some difficulties exist: 
All the corrections for inclination effects on both $B_{T}^{c}$ and $W$ depend on models
which use the axis ratio  $R_{25}$  of
the external isophotes and the morphological type code $T$ of the considered
galaxy. Further, it has been said that $a$ and $b$ also depend on $T$
(Roberts, 1978; Rubin et al., 1985; Theureau et al., 1997)
and it has been suggested that the linearity is not necessarily satisfied
(Aaronson and Mould, 1983; Mould and Han, 1989).
The use of the TF relation is plagued by these problems and the
resulting distance scale may appear less convincing.

The second problem is related to a statistical bias (the so-called Malmquist bias).
This bias can be understood as following. First of all, at large distances only
the intrinsically brightest galaxies are seen 
because any sample is limited to a given apparent magnitude.
In addition, the chance of finding galaxies over- or under-luminous (for
their $W$) is higher in a large sample, i.e. at a large distance.
These features are clearly illustrated by the Spaenhauer diagram (see Sandage, 1994), 
a diagram of absolute magnitude $M$ versus radial velocity $V$ (or distance).

In summary, if one wants to produce a sufficiently convincing value for the Hubble
constant three difficulties have to be solved:
\begin{itemize}
\item there must be a secure zero-point calibration
\item one has to correct for inclination effects and morphological type dependence.
\item we have to overcome statistical biases like the Malmquist bias.
\end{itemize}

The first item (zero-point) is very important and will be discussed in section 2
in the light of new results obtained in studying the catalog of HIPPARCOS Cepheids.
Then, the method of sosies-galaxies used to solve the second problem (corrections of
secondary effects) is discussed in section 3. 
The third item (Malmquist bias) is considered in section 4.
Finally, an application to sosie galaxies of M 31 and M 81 is made in section 5
and the results are presented in section 6.

\section{Discussion of the distance scale calibration}
From a general compilation of Cepheid measurements we derived
distances for 17 galaxies using geometrical parallaxes of 10
galactic Cepheids provided by the HIPPARCOS satellite (Paturel et al., 1997).
The result was obtained through a generalization of classical precepts.

Three important points should be emphasized:
\begin{itemize}
\item The equations were linear. Thus, they were used only on the range BVRI and only for
relatively small extinction (because the ratio of total to differential extinction is
assumed to be constant for a given effective wavelength).
\item The distance modulus of a given galaxy was directly connected to the mean distance modulus
$\overline{\mu}$ of the HIPPARCOS galactic Cepheid sample. Any change in $\overline{\mu}$ will affect
the final result. The uncertainty on $\overline{\mu}$ is large ($\sigma(\overline{\mu})\approx 0.2$)
because it is based mainly on 10 calibrating Cepheids (those having the highest weights).
\item Evidence is given that the Period-Luminosity relation (PL) is affected, at least in
some galaxies, by a statistical bias (incompleteness bias, similar to the Malmquist bias).
\end{itemize}

The distance moduli found in this way were about 0.2 magnitude larger than those
generally admitted. For instance, the distance modulus of Large Magellanic Cloud (LMC) was found 
to be $\mu(LMC)=18.7$ (Feast and Catchpole, 1997; Paturel et al., 1997), 
while RR Lyrae stars give $\mu(LMC)=18.3$ (Fernley et al., 1997), 
the SN1987 gives $\mu(LMC)=18.4 - 18.6$ (Gould, 1995; Panagia et al., 1997). From independent
studies on Cepheids, the distance moduli is found between 18.5 and 18.7 (Gieren et al., 1998; 
Gieren et al., 1993). Anyway, the distance modulus of LMC lies between 18.3 and 18.7. Our value
is in favour of the large ones.
In our first study (Paturel et al., 1997) the distance moduli for M 31 and M 81 
were $\mu(M31)=24.8$ and $\mu(M81)=28.1$ without
correction for the bias and $\mu(M31)=24.9$ and $\mu(M81)=28.2$, with a tentative correction.
Our result was in favour of a long distance scale. 
{\it Stricto Sensu}, our results are compatible with the generally admitted distance moduli
but the uncertainty on the zero-point is still large ($0.2$).

A new study has been undergone by one of us (PL) to revisit the HIPPARCOS PL calibration
after the release of the whole HIPPARCOS catalogue.  Some important conclusions are drawn:
\begin{itemize}
\item The value of $\overline{\mu}$ obtained from HIPPARCOS parallaxes 
must be based on confirmed solitary Cepheids.
We rejected binary stars in our original sample but many remaining ones may still be binaries.
This point was already underlined by Szabados (1997) and confirmed by us. 
A part of this effect can be understood by the nearness of confirmed solitary Cepheids which makes them
less sensitive to the Lutz and Kelker bias (1973). This does not change the practical conclusion:
It is better to base our zero-point on nearby solitary Cepheids.
\item Using all available solitary Cepheids from the HIPPARCOS mission we confirmed exactly the
V-band PL relation of Gieren and Fouqu\'e (1998) who give $\mu(LMC)=18.4$. Then, adopting their
I-band PL relation we applied the de-reddening method of Madore and Freedman (1991) and derived
$\mu(M31)=24.6$ and $\mu(M81)=27.6$, without correcting for the bias (i.e. with all periods).
\item The incompleteness bias on the Cepheid PL relation 
has been independently confirmed using available HST observations and numerical simulations. 
It is generally small (less than 0.2 magnitude)
because the dispersion of the PL relation is small. Nevertheless, correcting 
the bias (by constructing the
growth curve $\mu = f(logP)$) we derived $\mu(M31)=24.7$ and $\mu(M81)=27.7$.
\end{itemize}
In conclusion of this section, 
we will adopt $\mu(M31)=24.6\pm 0.2$ and $\mu(M81)=27.6\pm 0.2$ to derive the range of possible
Hubble constant but we will keep in mind that our study with HIPPARCOS catalogue 
leads to distance moduli $0.1$ mag larger
after correction for incompleteness bias on the Cepheids Period-Luminosity relation.

Now we have to discuss a method to avoid uncertainties in the correction of secondary effects
on the TF relation.
 
\section{Overcoming secondary effects with sosies}
We could make a model for correcting inclination effects and morphological 
type dependence. 
This necessarily leads to uncertainties because of the model dependency.
The way we have chosen here is different. We will not try to correct for
these perturbing effects but we will create a situation where
the correcting terms disappear.

If one selects galaxies having very nearly the
same morphological type, the same axis ratio and 
the same rotational velocity
than a given calibrating galaxy (here, M 31 or M 81), they will have the
same absolute magnitude as the calibrator. This is the direct
consequence of Rel.~1. This approach (Paturel, 1981) is called the
method of sosie galaxies  ("sosies" is the French word for
look-alike). Indeed, such galaxies will have the same
appearance and the same physical properties.
The distance modulus of a sosie galaxy is simply obtained
by the relation:
\begin{equation}
\mu = \mu (calib) + B_{T}^{c} - B_{T}^{c} (calib)
\end{equation}
where "calib" refers to the considered calibrator (M 31 or M 81) and $B_{T}^{c}$
is the apparent magnitude corrected for secondary effects.
In this case, the corrections for inclination effects and morphological 
type dependence have no incidence of the final result because inclination and morphological
type are the same. Nevertheless, a differential correction is required for
the galactic extinction, but it is quite small because this is
a differential secondary effect. This is still more valid because we
restrict the selection to low extinction regions.
Note that Sandage (1996) also made use of the concept of sosie galaxies
by considering galaxies having the same morphological type and the same
luminosity class. 

\section{Statistical bias}
In our first application of the method of sosies (Paturel, 1981)
we naively claimed that "this method is free from any Malmquist
bias" because we thought that at a given rotational velocity only
one single value of the absolute magnitude may exist.
In fact, this is not correct. Sosies galaxies do have
a dispersion of absolute magnitudes around a constant mean value. 
Making a magnitude limited sample selects
the brightest galaxies of the distribution and then biases the sample.

The importance of the bias in connection with morphological luminosity classes
was described as far back as 1975 (Sandage and Tamman, 1975; Teerikorpi, 1975a,b).
Two diagrams help one to understand the bias.
The {\it Plateau diagram} from which the method of the normalized distance is derived
(Teerikorpi, 1984, Bottinelli et al., 1986) and the {\it Spaenhauer diagram}
(Spaenhauer, 1978; Sandage, 1994).
The connection between these diagrams has been recently reviewed by Teerikorpi (1997)
who suggests the term {\it Malmquist bias of the second kind}
for the distance dependent bias, to distinguish it from the classical Malmquist bias.

\begin{figure}
\epsfxsize=8.5cm
\hbox{\epsfbox{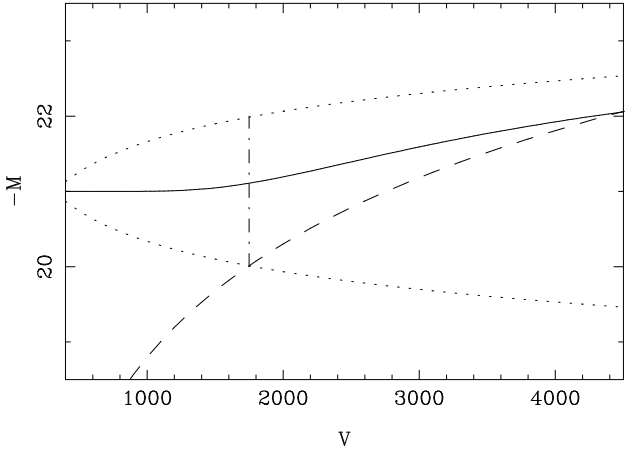}}
\caption{Spaenhauer diagram: for a sample of galaxies, limited to an apparent
magnitude and having on the mean the same
absolute magnitude only galaxies above the completeness limit are visible
(dashed curve). The dispersion around the mean increases with the size of the sample
(i.e. with the distance) but reaches a physical limit as shown by the
envelopes (dotted curves). The actual mean apparent absolute magnitude will follow the solid line.
The vertical line shows where the bias begins.}
\label{spaenhauer1}
\end{figure}

The Spaenhauer diagram shows how the absolute magnitude behaves as a function of the kinematical 
distance (i.e., distance estimated by the velocity). An illustration of this diagram is given
in Fig. \ref{spaenhauer1} for galaxies of a nominal absolute magnitude of $-21$.
The equation of the envelope of the distribution (dotted curves in Fig. \ref{spaenhauer1}) 
can be calculated as $|M'-M_o| = c.logV + d$, where $c= 2 \sigma /1.57$  and where $d$ is
a function of the space density of considered galaxies (Paturel, 1998).
$\sigma$ is the dispersion of the TF relation for the given rotation velocity.

The Plateau diagram shows how $logH$ changes with the kinematical distance. 
For galaxies assumed to have the same absolute magnitude (like the sosies of a given calibrator)
the Plateau diagram exhibits a plateau where $logH$ is unbiased (see Figures \ref{logHM31} and
\ref{logHM81}).
The method of the Normalized Distance 
consists in compressing or expanding the $x-$axis of the Plateau diagram
according to the absolute magnitude so that the plateau becomes the same whatever
the absolute magnitude of galaxies. This is achieved
by multiplying the 'distance' $V$ by $10^{0.2M}$. Because $M$ is not known {\it a priori}
it can be replaced through the relation by $10^{a/5(logV_{M})}$, 
where $a$ is the slope of the Tully-Fisher relation. 
\footnote{In the case of sosie galaxies we will use the absolute magnitude $M(calib)$ 
of the calibrator to perform the normalization}
The thus corrected distance is called the normalized distance. An example of the
normalized diagram is given in Figure \ref{logHN}.

The analytical curve for the Malmquist bias of the 2nd kind (Teerikorpi, 1975b) allows one
to predict the change of the mean absolute magnitude $\Delta M =M' - M_o$ 
(or the change $\Delta logH=logH' -logH_o$) with the 'distance' ($V$). $M'$ and $logH'$
refer to biased quantities. We have:
\begin{equation}
\Delta M= M' - M_o = \sigma {\sqrt{\frac{2}{\pi}} \frac {e^{-A^2}}{1+erf(A)}}
\end{equation}

with
\begin{equation}
A= \frac {M_{lim} - M_o} {\sigma \sqrt{2}}
\end{equation}
and
\begin{equation}
erf(x) = \frac{2}{\sqrt{\pi}} \int_0^x e^{-t^2} dt
\end{equation}
and 
\begin{equation}
\label{eqMlim}
M_{lim} = m_{lim} -25 -5logV/H_o 
\end{equation}

The bias curve in the Plateau diagram is simply: 
\begin{equation}
\Delta logH = logH' - logH_o = -0.2 \Delta M
\label{eqbias}
\end{equation}

The completeness limit  curve (dashed curve in our diagrams) will show where the
sample is cut. No galaxies will appear below this limit for the considered absolute magnitude.
In the Spaenhauer diagram the completeness limit is given by Eq. \ref{eqMlim}. This curve is
unique whatever the absolute magnitude (the equation does not depends on $M_o$).
In the Plateau diagram the limit can be deduced from:
\begin{equation}
logH_{lim}=0.2 M_o +5 -0.2 m_{lim} + logV
\end{equation}
In the Plateau diagram it is obvious that there is a completeness limit for
each calibrator because $logH_{lim}$ explicitly depends on $M_o$. 
When the normalization is applied on $x$-axis multiplying
by $10^{0.2M_o}$, the completeness limit becomes unique for each $M_o$.

\section{Application to M31 and M81 sosie galaxies}

\subsection{The sample}
In the present paper we use only the calibrators M31 and M81 because 
they are the brightest ones. The main characteristics of these galaxies
are given in Table 1. All observed astrophysical parameters are taken from the
LEDA database which is freely accessible 
\footnote{{\bf telnet:}   telnet leda.univ-lyon1.fr,   login: leda\\
or {\bf world-wide-web:} http:www-obs.univ-lyon1.fr/leda}.
Distance moduli are taken according to the discussion of section 2 and the
absolute magnitudes are derived from them using the corrected apparent
magnitudes. Note that the correction for galactic extinction on magnitudes
are taken following the RC2 system (de Vaucouleurs et al., 1976). 
The essential difference with the Burstein-Heiles
model occurs near the galactic poles (see Paturel et al., 1997) 
where the extinction is assumed to be 0.2 magnitudes. However, 
this has no influence on the present results because the constant term 
is cancelled in our differential method.
   
\begin{table}
\begin{tabular}{|l|r|r|}
\hline
parameter      &  M 31                   &    M 81                 \\
\hline
$\mu$          & $24.6 \pm 0.2$         &  $27.6 \pm 0.2$        \\
$M_B$          &$-21.49           $     & $-20.62           $    \\
$T code$       &  $3.0   \pm 0.3  $     &  $ 2.5   \pm 0.6  $    \\
$logR_{25}$    &  $0.48  \pm 0.05 $     &  $ 0.31  \pm 0.07 $    \\
$B_{T}^{c}$    &  $3.11  \pm 0.4  $     &  $ 6.98  \pm 0.32 $    \\
$logV_M$       &  $2.397 \pm 0.010$     &  $ 2.338 \pm 0.022$    \\
\hline
\end{tabular}
\caption{Parameters of Calibrating Galaxies}
\end{table}

Using the LEDA database we selected
sosie galaxies, i.e. galaxies having nearly the same morphological type code ($T$),
the same log of axis ratio ($logR_{25}$) and the same rotational velocity ($logV_M$), 
as M 31 or M 81. Further, we selected only galaxies with 
accurate apparent magnitude ($ \sigma(B_T) \leq 0.4 $) and 
rotational velocity ($ \sigma(logV_M) \leq 0.08 $) and no discrepancies in
observed radial velocity.
The tolerances are chosen to equal a typical error in each parameter. They are
$\pm 1.5$ for the morphological type code, $\pm 0.12$ for
$logR_{25}$ and $\pm0.08$ for $logV_M$.

Using such criteria we obtained a sample of 43 sosies
of M 31 and 119 sosies of M 81. 
For each galaxy the following parameters are known
(they all conform to the description given by Paturel et al., 1997):

\begin{itemize}
\item Column 1: PGC name according to Paturel et al. (1989a,b)
\item Column 2: Right ascension and Declination for 2000 in (h,mn,s) and
(deg,arcmin,arcsec) respectively.
\item Column 3: Corrected total $B_{T}^{c}$ magnitude in the RC3 system 
(de Vaucouleurs et al., 1991) - except for the galactic extinction. 
\item Column 4: Radial velocity corrected for the infall of the Local Group onto
the Virgo cluster. The infall velocity is $170$ $km.s^{-1}$ according to Sandage 
and Tammann (1990), the position of the Virgo cluster is $SGL=104 \deg$
and $SGBD=-2 \deg$, in supergalactic coordinates. This velocity is
used as a relative kinematical distance.
\end{itemize}

As we saw in previous section, in order to use the Normalized Distance method
it is compulsory to have a sharp apparent limiting
magnitude in our sample. 
This magnitude limit and the resulting useful sample
can be derived by constructing the curve $logN$ vs. the
apparent magnitude $B_{T}^{c}$ ($N$ is the cumulative 
number of galaxies with $B_{T}^{c}<B_{lim}$). 
The result must be a straight line. The magnitude at
which the curve bends down gives  the limiting magnitude $B_{lim}$.
The slope should be $0.6=3/5$ if the distribution of galaxies is homogeneous
($N \propto r^3$). In practice, this value is rarely reached. The slope is
always lower. In the local universe it has been shown that the slope is
close to $0.4=2/5$ (Paturel et al., 1994). This is also consistent
with a recent derivation of the radial space density by a method using
the Tully-Fisher distance moduli (Teerikorpi et al., 1998).

\begin{figure}
\epsfxsize=8.5cm
\hbox{\epsfbox{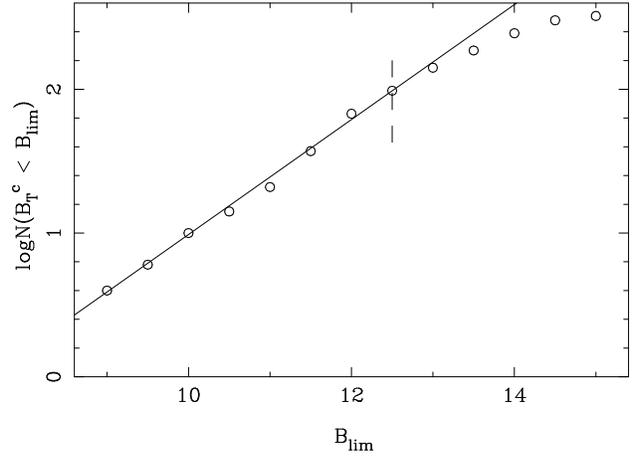}}
\caption{Completeness curve of the sample. This curve is used to
define the apparent limiting magnitude of our sample in an objective
manner. The completeness is fulfilled up to an apparent magnitude of 
$B_{lim} \approx 12.5$ (vertical line).}
\label{complet}
\end{figure}

For our sample the completeness curve $logN$ vs $B_{T}^{c}$ is built 
(Fig. \ref{complet}) by combining sosies of both calibrators M31 and M81 (the apparent
limiting magnitude is an observational limit which does not depend on the
absolute magnitude of considered objects). The completeness is
fulfilled up to an apparent magnitude of $B_{lim} \approx 12.5$.
The slope is in agreement with our previous result ($0.4$).

Thus, the diagrams will be made using only galaxies brighter than
$B_{T}^{c} = 12.5$. The corresponding sample is given in the Annex.

\subsection{Construction of the Plateau diagram}
\begin{figure}
\epsfxsize=8.5cm
\hbox{\epsfbox{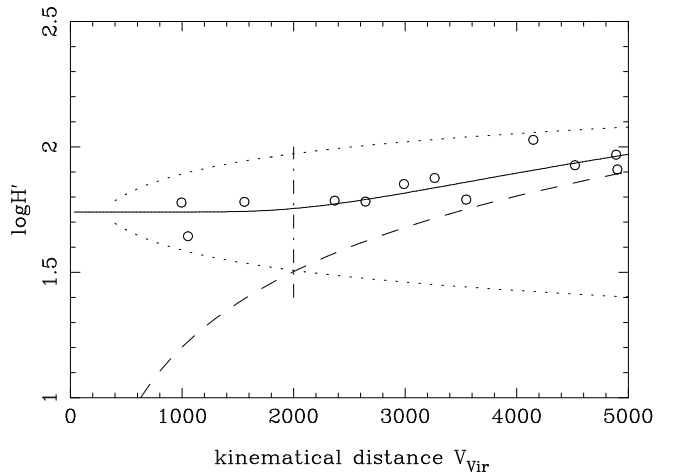}}
\caption{ The Plateau diagram ($logH'$ vs. $V_{Vir}$) for sosie galaxies of M 31. 
(the velocity is used as an estimator of relative distances).
As expected, $logH'$ increases with the distance because of the Malmquist bias.
The full line represents the model of the bias from Teerikorpi (1975b). 
The dashed line represents
the limit in apparent magntiude ($B_{lim}=12.5$). The dotted curves represent the
envelope of the distribution.  The vertical line corresponds to the
distance where the bias begins. The adopted $H_o$ is 50 $km.s^{-1}.Mpc^{-1}$.
}
\label{logHM31}
\end{figure}

\begin{figure}
\epsfxsize=8.5cm
\hbox{\epsfbox{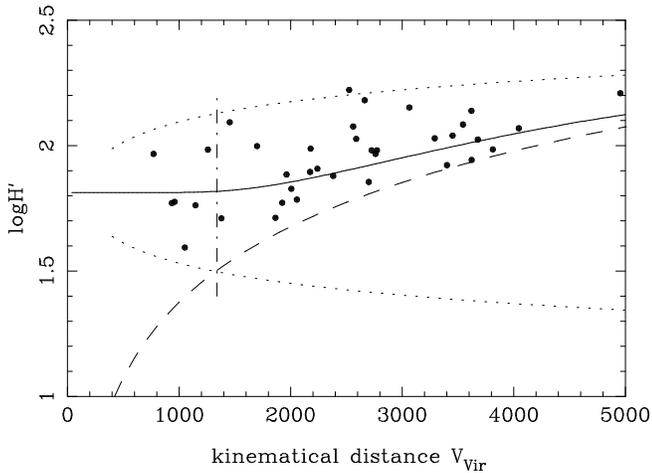}}
\caption{ The Plateau diagram for sosie galaxies of M 81.
(same comments as the previous figure). The adopted
$H_o$ is 65 $km.s^{-1}.Mpc^{-1}$.}
\label{logHM81}
\end{figure}

The Hubble constant is calculated from the relation:

\begin{equation}
logH' = logV_{Vir} -0.2[\mu (calib) + B_{T}^{c} - B_{T}^{c} (calib) -25]
\end{equation}

for each galaxy for which $V_{Vir} \geq 500 km.s^{-1}$ 
(with this radial velocity limit we avoid the effect of random 
peculiar velocities).
Figures \ref{logHM31} and \ref{logHM81} show
the Plateau diagrams, $logH'$ vs. $V_{Vir}$, for M31 and M81 respectively. 
The bias curve (Eq.\ref{eqbias}) is calculated
using $B_{lim}=12.5$ (previous section)  and $\sigma=0.7 mag$ (Fouqu\'e et al., 1990).
The expected increase of $logH'$ is clearly visible.
It reveals the presence of a Malmquist bias.
The velocity at which the bias becomes significant (i.e. $\Delta M \approx 0.05$)
is 2000 $km.s^{-1}$ for M 31 and 1400 $km.s^{-1}$ for M 81.

For sosie galaxies of a given calibrator, the Plateau diagram has the same meaning
than the Spaenhauer diagram. The deviation of $logH'$ simply reflects the
corresponding deviation of $M'$ in the Spaenhauer diagram.

\begin{figure}
\epsfxsize=8.5cm
\hbox{\epsfbox{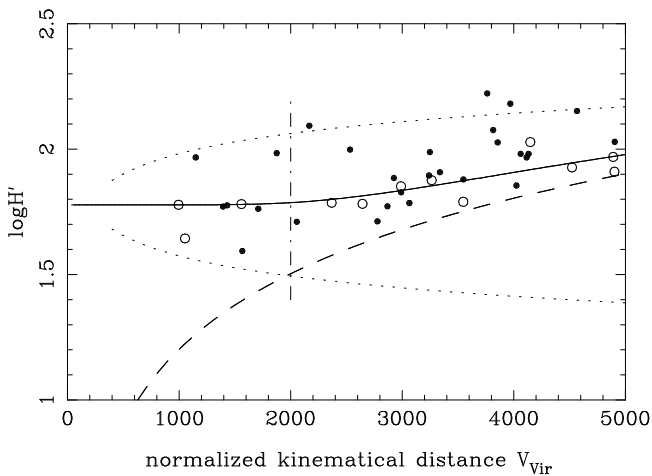}}
\caption{ The Normalized distance ($logH'$ vs. $V_{Vir}10^{-0.2(\Delta M_o)}$) 
for sosie galaxies of both M31 and M81 (see text for the calculation of the
normalization factor). The mean $H_o$ adopted is $60km.s^{-1}.Mpc^{-1}$.}
\label{logHN}
\end{figure}

We can draw the Normalized distance diagram by superimposing both Plateau diagrams after
having expanded the $x-$axis for M81 by the factor $10^{-0.2(M_o(M31)-M_o(M81))}$, i.e.
$1.49$. Note that in this case we adopted a mean Hubble constant $H_o=60km.s^{-1}.Mpc^{-1}$.
The result is given in Fig.\ref{logHN}. From this diagram it appears that the bias begins
near a normalized 'distance' of $V_{Vir}=2000km.s^{-1}$ 
(i.e., $1.49 \times 1400$)
This will limit the unbiased sample to only 9 galaxies which
will be used to define the most likely value of the Hubble constant $H_o$.

\subsection{Mean Hubble constant derived from the geometrical zero-point}
The final sample of unbiased galaxies is given in Table \ref{T-finale}.
It is rather small as expected from the size of the sosies sample and 
from the restriction $V_{Vir} \geq 500 km.s^{-1}$.
\begin{table}
\begin{tabular}{|l|r|r|r|}
\hline
M 31 sosies &$B_{T}^{c}$&$V_{Vir}$&$logH_o$\\
\hline
PGC14638 &  9.61 &  995  & 1.778  \\
PGC39724 & 10.40 & 1053  & 1.644  \\
PGC51233 & 10.57 & 1559  & 1.781  \\
\hline
M 81 sosies &$B_{T}^{c}$&$V_{Vir}$&$logH_o$\\
\hline
PGC00218 & 10.87 & 1147  & 1.762  \\
PGC10208 & 10.41 &  960  & 1.776  \\
PGC33550 &  8.98 &  771  & 1.967  \\
PGC35164 & 10.38 &  935  & 1.771  \\
PGC37306 &  9.96 & 1258  & 1.984  \\
PGC43798 & 11.52 & 1051  & 1.594  \\
\hline
mean     &      &       &$1.78\pm0.04$\\
\hline
\end{tabular}
\caption{Sample of unbiased galaxies}
\label{T-finale}
\end{table}

From the mean $logH_o=1.78\pm0.04$ we derive the Hubble constant
$H_o = 60\pm 5(internal)  km.s^{-1}.Mpc^{-1}$.
If we are using the extreme values for $\mu_{calib}$ we find that the
Hubble constant lies between $47$ and $65 km.s^{-1}.Mpc^{-1}$.

\section{Conclusion}

A reasonable value of the unbiased Hubble constant would be:
\begin{equation}
H_o = 60\pm 10 (external)  km.s^{-1}.Mpc^{-1}
\end{equation}

Nevertheless, we believe that the correct value is more probably near the
lower limit because we have demonstrated that the PL relation of Cepheids
also suffers from a statistical bias.

It is clear, that the bias appears already at a short distance, 
even when a luminous galaxy is used as calibrator. Hence, it is
difficult to obtain a secure Hubble constant from
low luminosity galaxies. Indeed, then restricting the sample to the
unbiased plateau requires a cut at a very low radial velocity,
just where random velocities start to dominate. This justifies that
we used first M 31 and M 81 as calibrators. However, one should ask
whether these galaxies have an absolute magnitude representative of
their rotation velocity. 
The positions of M 31 and M 81 in a TF diagram are quite normal (see
for instance Fouqu\'e et al., 1990), which supports that they 
are good calibrators.

The positive sides of the sosies method make it necessary to enlarge
the sample in the future.
This can be achieved only by using deeper samples which require
more HI measurements. Besides, weaker sosies criteria which require that
galaxies have the same morphological type and the same axis ratio,
but not necessarily the same 21-cm line width $W$ as 
a calibrator will greatly improve the accuracy of the TF relation
because inclination and morphological type effects will be
cancelled.

Another way to improve the result may come from infrared
photometry. Indeed, even with the method of sosie galaxies it is
necessary to correct for galactic extinction. Obviously, only
the difference between the extinctions in front of the sosies galaxy
and the calibrator enter the final result. 
Nevertheless, it is still a part of the uncertainty.

In view of this promising aspects,
we are engaged in a program aiming at collecting deeper samples
of sosie galaxies especially in the infrared domain.

\acknowledgements{We express our gratitude to A.R. Sandage for 
valuable comments on the first version of this paper. We want
also to thank C. Petit for her helpful contribution during the preparation 
of this paper. P.T. acknowledges the support from the Academy of Finland
(project "Cosmology in the local galaxy Universe").
}

\newpage
\begin{verbatim}
Annex: Samples of M 31 and M 81 sosie galaxies 
brighter than BTc=12.5.

-----------------------------------------
PGC name  RA.(2000)DEC.      BTc   V(Vir)
=========================================
M 31      004244.4+411608    3.11    -116
PGC05268  012519.7+340128   12.41    4902
PGC05344  012621.6+344214   12.22    5338
PGC10048  023912.2+105050   12.31    3547
PGC14638  041205.5-325228    9.61     995
PGC26157  091619.6-233804   11.45    2368
PGC36964  114950.1-384705   11.71    2644
PGC39724  121950.9+293654   10.40    1053
PGC51233  142020.5+035559   10.57    1559
PGC54001  150735.4+193457   12.11    4891
PGC55256  153007.8-383857   12.15    4522
PGC55740  153958.1+580457   11.70    3264
PGC62951  191443.8-621618   11.46    4148
PGC63214  192649.8-545704   11.63    2989
PGC63464  193842.8-294832   12.31    6011
-----------------------------------------
\end{verbatim}
\newpage
\begin{verbatim}

-----------------------------------------
PGC name  RA.(2000)DEC.      BTc   V(Vir)
=========================================
M 81      095533.5+690400    6.98     201
PGC00218  000315.1+160845   10.87    1147
PGC03260  005507.9+313229   12.22    5606
PGC04777  011945.5+032437   11.87    2381
PGC07525  015920.3+190022   10.28    2524
PGC10208  024144.7+002631   10.41     960
PGC13059  033108.4-333744   10.54    1699
PGC13108  033203.1-204904   11.53    1379
PGC13179  033336.6-360817    9.73    1454
PGC13584  034157.2-044221   12.07    4045
PGC18355  060340.2-693541   12.09    3678
PGC23086  081414.4+212125   12.43    3401
PGC23362  081948.2+220138   12.46    3622
PGC24427  084135.0-201858   11.76    5555
PGC24464  084240.1+141710   12.02    2056
PGC24558  084430.1-202101   11.82    3291
PGC25097  085607.9-032139   11.94    1924
PGC28376  095115.8-324518   11.31    2587
PGC29591  101009.9-122602   11.71    3545
PGC29993  101619.3-333353   11.65    2725
PGC31335  103508.5-434132   10.60    2663
PGC31533  103715.6-413742   11.04    2560
PGC33550  110548.9-000215    8.98     771
PGC35164  112607.7+433506   10.38     935
PGC37306  115349.5+521939    9.96    1258
PGC39212  121533.7-353746   11.69    2773
PGC39656  121922.0+060601   12.17    1864
PGC43798  125329.1+021011   11.52    1051
PGC45170  130414.6-102021   11.05    3064
PGC46304  131743.4-320605   11.59    2174
PGC49359  135328.1+375452   12.36    3812
PGC49563  135616.8+471417   11.75    2006
PGC50031  140250.5+491025   11.59    2240
PGC51932  143205.2+575524   11.13    2179
PGC54364  151345.7-141615   11.42    1963
PGC62178  183833.7+252238   11.48    3620
PGC62528  185255.5-591520   11.87    3451
PGC64601  202334.0+062638   11.81    4955
PGC64884  203138.5-305001   11.75    2761
PGC67839  220102.0-131610   12.26    2700
-----------------------------------------
\end{verbatim}

\end{document}